\documentclass[prc,twocolumn,twoside,floatfix,nofootinbib]{revtex4}

\usepackage{amssymb}
\usepackage{graphicx}
\usepackage{dcolumn}
\usepackage{bm}
\usepackage{float}
\usepackage{color}
\usepackage{mathtools}
\usepackage{scrextend}
\usepackage{tabularx}
\usepackage{booktabs}
\usepackage[normalem]{ulem} 
\emergencystretch=\maxdimen
\begin{document}

\title{Full realization of the RIBLL2 separator at the HIRFL-CSR facility}
\author{Xiao-Dong Xu$^{1,2}$}
\author{Yong Zheng$^{1}$}\thanks{Corresponding author: zhengyong@impcas.ac.cn}
\author{Zhi-Yu Sun$^{1,2}$}\thanks{Corresponding author: sunzhy@impcas.ac.cn}
\author{Yu-Nan Song$^{1,2}$}
\author{Bao-Hua Sun$^{3}$}\thanks{Corresponding author: bhsun@buaa.edu.cn}
\author{Satoru Terashima$^{1}$}
\author{Chang-Jian Wang$^{3}$}
\author{Ge Guo$^{3}$}
\author{Guang-Shuai Li$^{3}$}
\author{Xiu-Lin Wei$^{3}$}
\author{Jun-Yao Xu$^{3}$}
\author{Ji-Chao Zhang$^{3}$}
\author{Yong Cao$^{3}$}
\author{Bing-Shui Gao$^{1,2}$}
\author{Jia-Xing Han$^{3}$}
\author{Jin-Rong Liu$^{3}$}
\author{Chen-Gui Lu$^{1}$}
\author{Shu-Ya Jin$^{1}$}
\author{Hooi Jin Ong$^{1,2}$}
\author{Hao-Tian Qi$^{3}$}
\author{Yun Qin$^{3}$}
\author{Ya-Zhou Sun$^{1}$}
\author{Isao Tanihata$^{3}$}
\author{Lu-Ping Wan$^{3}$}
\author{Kai-Long Wang$^{1}$}
\author{Shi-Tao Wang$^{1,2}$}
\author{Xin-Xu Wang$^{3}$}
\author{Tian-Yu Wu$^{3}$}
\author{Xiao-Tian Wu$^{1}$}
\author{Mei-Xue Zhang$^{3}$}
\author{Wen-Wen Zhang$^{3}$}
\author{Xiao-Bin Zhang$^{3}$}
\author{Xue-Heng Zhang$^{1,2}$}
\author{Jian-Wei Zhao$^{3}$}
\author{Zi-Cheng Zhou$^{3}$}
\author{Li-Hua Zhu$^{3}$}

\affiliation{$^1$Institute of Modern Physics, Chinese Academy of Sciences, Lanzhou 730000, China}
\affiliation{$^2$School of Nuclear Science and Technology, University of Chinese Academy of Sciences, Beijing 100049, China}
\affiliation{$^3$School of Physics, Beihang University, Beijing 100191, People’s Republic of China}

\date{\today}

\maketitle

Since the pioneering use of radioactive ion beams (RIB) in the 1980s at Lawrence Berkeley National Laboratory in USA, production and investigation of exotic nuclei far from stability have become possible for a wide range of nuclear chart.~Over the past few decades, the RIB has triggered tremendous advancements in the study of exotic nuclei. For instance, over 3300 nuclides have been discovered~\cite{Kondev2021}, with a substantial number of them produced by employing RIBs.~Experimentally, there are several methods to generate radioactive nuclei. Among them, the in-flight production methods, specifically projectile fragmentation and in-flight fission, are proven to be two of the most crucial approaches. In-flight methods have been widely used to create the radioactive nuclei lying away from the valley of stability, such as those situated on or close to the \emph{r}-process path (e.g., see Refs.~\cite{Bernas1997,Ohnishi2010,Nieto2014}), thereby allowing scientists to explore the previously inaccessible nuclear territory.~In the past few decades, a number of fragment separators have been constructed worldwide in order to produce exotic nuclei utilizing the in-flight techniques~\cite{Ma2021}. Prominent setups include FRS at GSI (Germany), LISE at GANIL (France), RIPS and BigRIPS at RIKEN (Japan), A1200 and A1900 at NSCL/MSU (USA), ARIS at FRIB/MSU (USA), ACCULINNA at DUBNA (Russia), RIBLL1 and RIBLL2 at HIRFL (China), etc.

Among the aforementioned experimental facilities, the Heavy Ion Research Facility in Lanzhou (HIRFL) is a multi-purpose accelerator complex which can provide beams of all ion species, from hydrogen to uranium. After an upgrade in the early 21st century, HIRFL was extended to the HIRFL-CSR facility, which consists of two cyclotrons, one synchrotron Cooler Storage main Ring (CSRm), and a Cooler Storage experimental Ring (CSRe)~\cite{Xia2002,Zhan2008}.~As a transit beam line connecting CSRm and CSRe, the Second Radioactive Ion Beam Line in Lanzhou (RIBLL2)~\cite{Xia2002,Song2001} is one of the few fragment separators around the world that can provide the RIBs with energies higher than 300 MeV/u.~Here we present several basic parameters of the RIBLL2 separator.~The maximum magnetic rigidity (B$\rho$) is 10.64 Tm.~The horizontal and vertical angular acceptances are $\pm 25$ mrad, while the momentum acceptance is $\pm 1$\%.~The horizontal and vertical magnifications from F0 to F1 are -0.487 and -11.77, respectively.~The momentum dispersion at dispersive focus F1 is 1.169 cm/\%.~The first order momentum resolving power at F1 is 1200, assuming that the object size at F0 has the half width of 1 mm.~A comprehensive description of the RIBLL2 can be found in Ref.~\cite{Song2001}.

\begin{figure}[!htbp]
\centerline{\includegraphics[scale=0.56, angle=0]{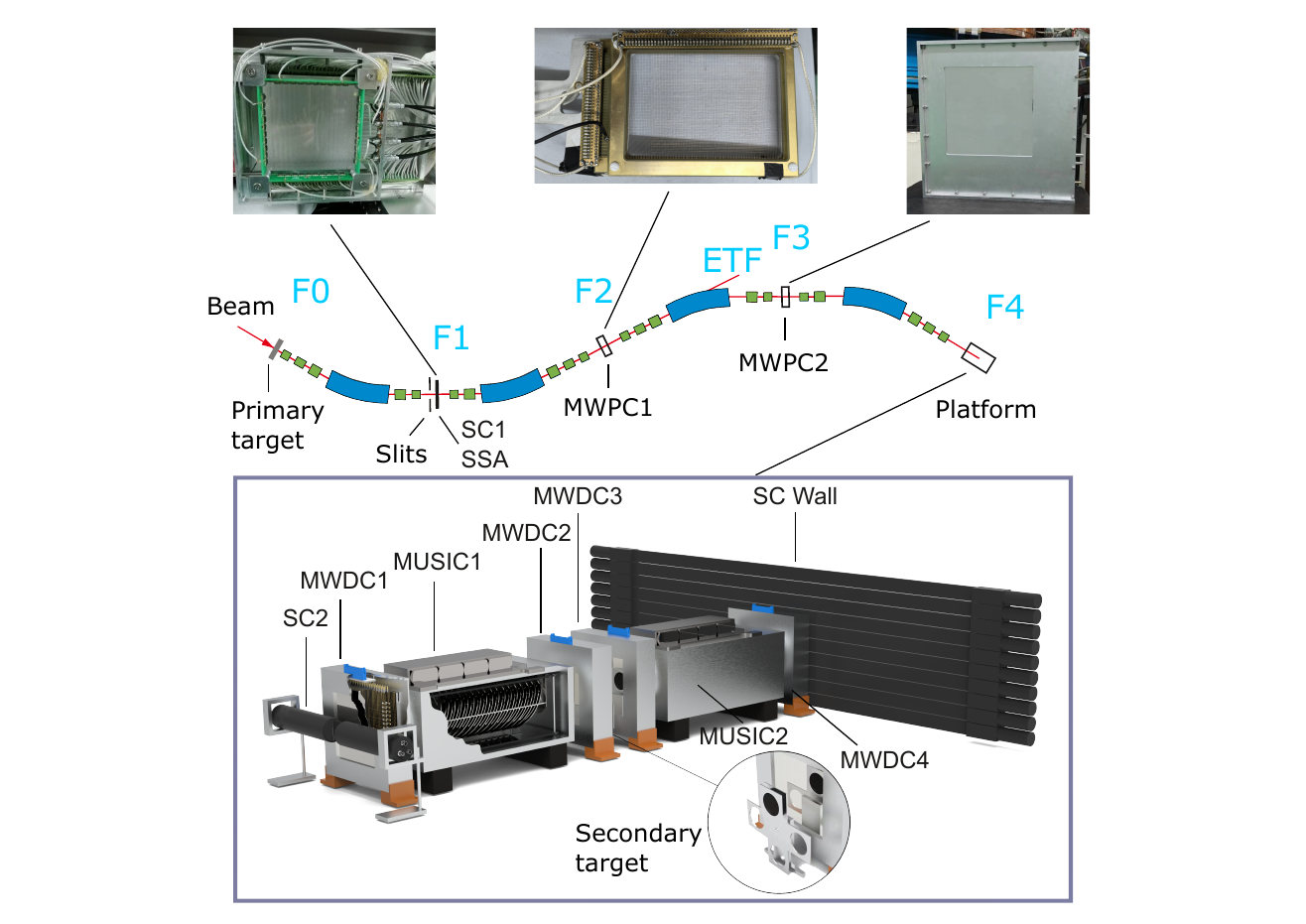}}
\caption{Schematic layout of RIBLL2 separator. Primary beam was injected into the entrance of the RIBLL2 to impinge on the primary target located at F0.~At the focal planes F1, F2, and F3, several beam line detectors (some of them are shown by photos) and instruments (such as the slits) are positioned.~At F4, the RIBLL2-F4 platform encompassing various detectors and the secondary target is illustrated by the lower schematic drawing. See text for details.}
\label{fig1}
\end{figure}

In the RIBLL2 beam line, an experimental platform called External Target Facility (ETF) was constructed after the RIBLL2 has been built.~It is situated downstream of the middle focal plane (F2) of the RIBLL2 (see Fig.~\ref{fig1}).~Typically, a secondary target is installed at ETF for reaction studies.~The ETF has a dipole magnet and various detectors, which allows an identification of RIBs transmitted from the RIBLL2 and a kinematics measurement of nuclear reactions occurred at the secondary target~\cite{Sun2019NIMA}.~By producing secondary beams at F0 using projectile fragmentation method, and then following that with the separation and transportation of nuclei of interest via beam line from F0 to F2, a series of experiments has been carried out at ETF, such as knock-out reaction studies (e.g., Ref.~\cite{Sun2014PRC}), production cross section measurements (e.g., Ref.~\cite{Xu2022}), charge-changing cross section (CCCS) measurements (e.g., Refs.~\cite{Zhao2023,Li2024}), and charge radius investigations~\cite{Zhao2024}~based on the recently discovered robust correlation between CCCS and separation energies~\cite{Zhang2024}.~Since the beam line from F0 to F2 has a limited separation capability and relatively short flight path ($\sim$26 m), the aforementioned experiments were mainly dedicated to study the \emph{p-} or \emph{sd-} shell nuclei by employing the light primary beams (such as $^{18}$O and $^{40}$Ar).~To further expand research on exotic nuclei toward medium- and heavy-mass regions, it is essential to fully realize the capabilities of the RIBLL2 separator~\cite{Sun2018}. The main task was to construct a new experimental platform at the last focal plane (F4) area thus the entire RIBLL2 beam line (55 m) could be fully utilized to separate and transmit the RIBs. Correspondingly, various detectors were developed for this new experimental platform (hereafter RIBLL2-F4 platform).~At the same time, some of the existing detectors were upgraded with the newly constructed one. This article summarizes these developments and introduces the current status of the RIBLL2 separator. Figure~\ref{fig1} is a schematic layout of RIBLL2, in which the beam line detectors and instruments currently installed at various focal planes together with the newly-developed RIBLL2-F4 platform are shown.
%-------------------------------------------------------------------------------
%-------------------------------------------------------------------------------

At the focal plane F1, a novel detector system was developed to replace the original plastic scintillation detector.~This system includes a plastic scintillation sheet (SC1 in Fig.~\ref{fig1}) and a plastic scintillator strip array (SSA in Fig.~\ref{fig1}) for measuring the timing and position of the ions. The active area of the SC1 is $100\times100~\rm{mm}^2$, while the SSA consists of 50 strips, each with a width of 2 mm, enabling the measurement of ion's coordinate in the horizontal direction (i.e., transverse coordinate relative to the beam direction). This allows a momentum resolution of around $1.9\times10^{-3}$ assuming an object size with the half width of 1 mm at F0.~The performance of this detector system has been demonstrated through the measurement of the time-of-flight (TOF) and the position at F1 for reaction products produced from the fragmentation of an energetic $^{78}$Kr beam.~The particle identification (PID) capability has been considerably enhanced by incorporating ion’s position measurement at F1~\cite{Fang2022}.

At the F2 area, a compact device housing a Multi-Wire Proportional Chamber (MWPC1 in Fig.~\ref{fig1}) has been specifically implemented for the position measurements.~This device is conveniently designed for insertion and removal from the vacuum chamber at F2. The MWPC1 has an active detection area of $100\times70~\rm{mm}^2$. It allows us to measure the position of particles at F2. This measurement serves a dual purpose. Firstly, it facilitates the monitoring of the position of beam spot, which in turn assists in the beam tuning process, particularly during the fine-tuning of the secondary beams. Secondly, a precise measurement of ion’s position at F2 is necessary for deducing the ion's magnetic rigidity.

At the F3 area, originally there was solely beampipe without any detectors. We removed a part of beampipe in order to create space for the installation of detectors in the air. A MWPC (MWPC2 in Fig.~\ref{fig1}) with an active area of $256\times256~\rm{mm}^2$ was developed and installed at F3. The size of MWPC2 is sufficiently large to accommodate the beam spot size at F3.

At the last focal plane F4, we redesigned the beampipe and constructed a motor-controlled platform to accommodate the installation of detectors. Subsequently, effort was devoted to the detector development.~A plastic scintillator coupled with photomultipliers was developed for timing measurement at F4.~The active area of this detector is $50\times50~\rm{mm}^2$. Four Multi-Wire Drift Chambers (MWDC) were constructed for the position measurement and trajectory reconstruction.~The effective volume of the MWDC is $160\times160\times70$ mm$^3$. Two MUlti-Sampling Ionization Chambers (MUSIC) were developed for ion’s charge measurement.~The MUSIC chamber has a total length of 520 mm, with 448 mm for housing the working gas, while its window has a diameter of 90 mm. Furthermore, a detector array consisting of 10 elongated shape scintillator detectors was built to detect the light charged particles. All the above-mentioned detectors were installed at the RIBLL2-F4 platform and their relative locations can be seen in Fig.~\ref{fig1}. The secondary target is placed in the middle of the platform. Upstream of the secondary target, a plastic scintillation detector (SC2 in Fig.~\ref{fig1}) and a MUSIC detector (i.e., MUSIC1 in Fig.~\ref{fig1}) sandwiched by two MWDCs (MWDC1 and MWDC2 in Fig.~\ref{fig1}) are placed. Downstream of the secondary target, another MUSIC detector (MUSIC2 in Fig.~\ref{fig1}) and two MWDCs (MWDC3 and MWDC4 in Fig.~\ref{fig1}) are installed. The scintillator detector array (SC Wall in Fig.~\ref{fig1}) is positioned at the farthest end of the platform. Such a detector configuration allows a PID before the reaction target and a charge identification after the reaction target, which are well suited for the CCCS measurements. It is worth mentioning that it is also feasible to substitute the MWDCs and MUSICs with silicon detectors to create additional room for detector arrays near the secondary target.

After accomplish of the RIBLL2-F4 experimental platform, an experiment aimed to measure the CCCS of various \emph{sd}-shell nuclei was carried out. A primary $^{40}$Ar beam was accelerated to 400 MeV/u and injected into the entrance of RIBLL2 where a 10 mm-thick Be target was located. The RIBLL2 was tuned to transport the RIBs of interest down to F4 to bombard a 10 mm-thick C target. Most of the detectors depicted in Fig.~\ref{fig1} were employed.~The TOF of ions was measured by using the SC1 and the SC2.~The MWPC2 was utilized to monitor the ion’s position, which was employed to gain insights into its momentum value. The MUSIC1 was employed to measure the energy loss ($\Delta{E}$) of ions upstream of the C target. The MWDC1 and MWDC2 were used to track the incoming particles. Downstream of the C target, the charge of outgoing ions was measured by the MUSIC2, while their trajectories were monitored by the MWDC3 and MWDC4. The working gas in the MUSICs was CF4 at 1 atm, while the MWDCs used P10 at 1 atm. Regarding the MWPCs, the working gas was a mixture of Ar (80\%)+$\mathrm{CO}_{2}$ (20\%) at 1 atm. In the following, the first results obtained in this experiment are presented to demonstrate the performance of the RIBLL2-F4 platform.

%-------------------------------------------------------------------------------
\begin{figure}[!htbp]
\centering
\begin{minipage}[b]{0.95\linewidth}
\centering
\includegraphics[scale=0.4, angle=0]{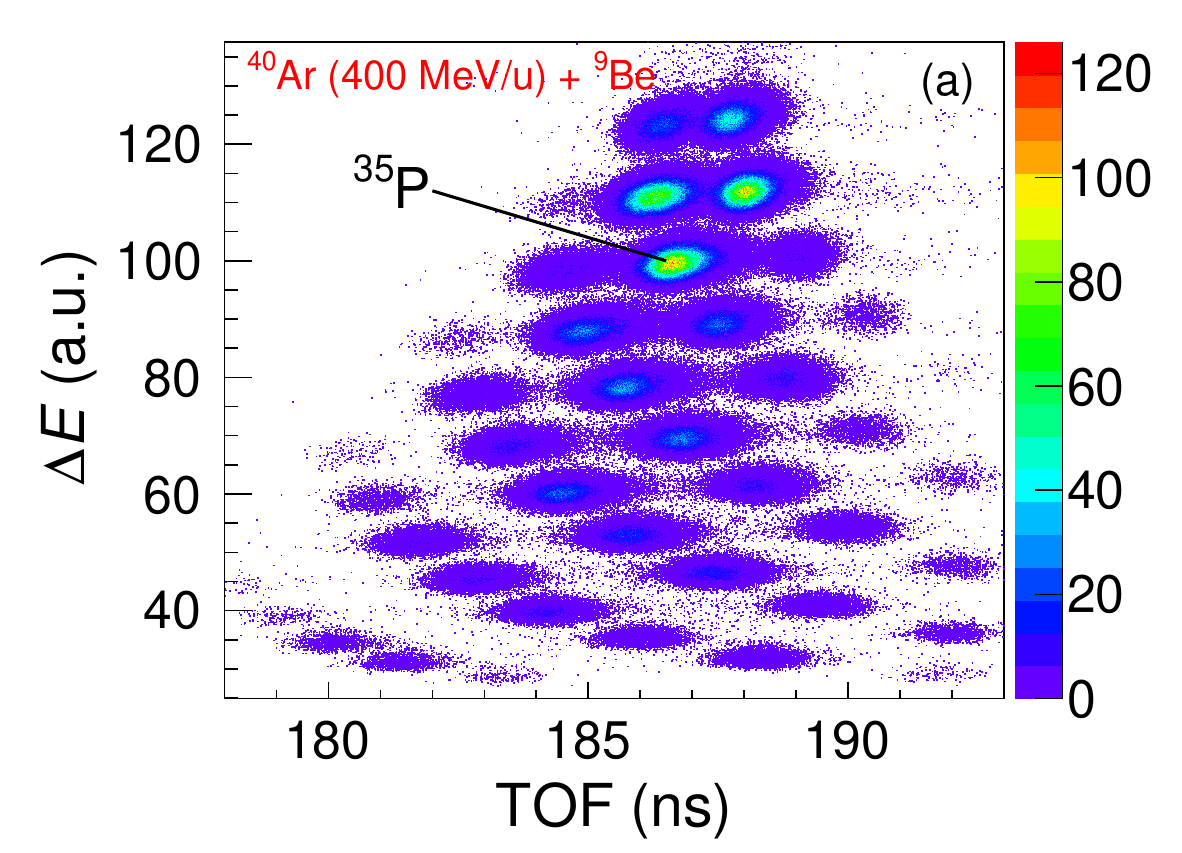}
\end{minipage}
\begin{minipage}[b]{0.95\linewidth}
\centering
\includegraphics[scale=0.4, angle=0]{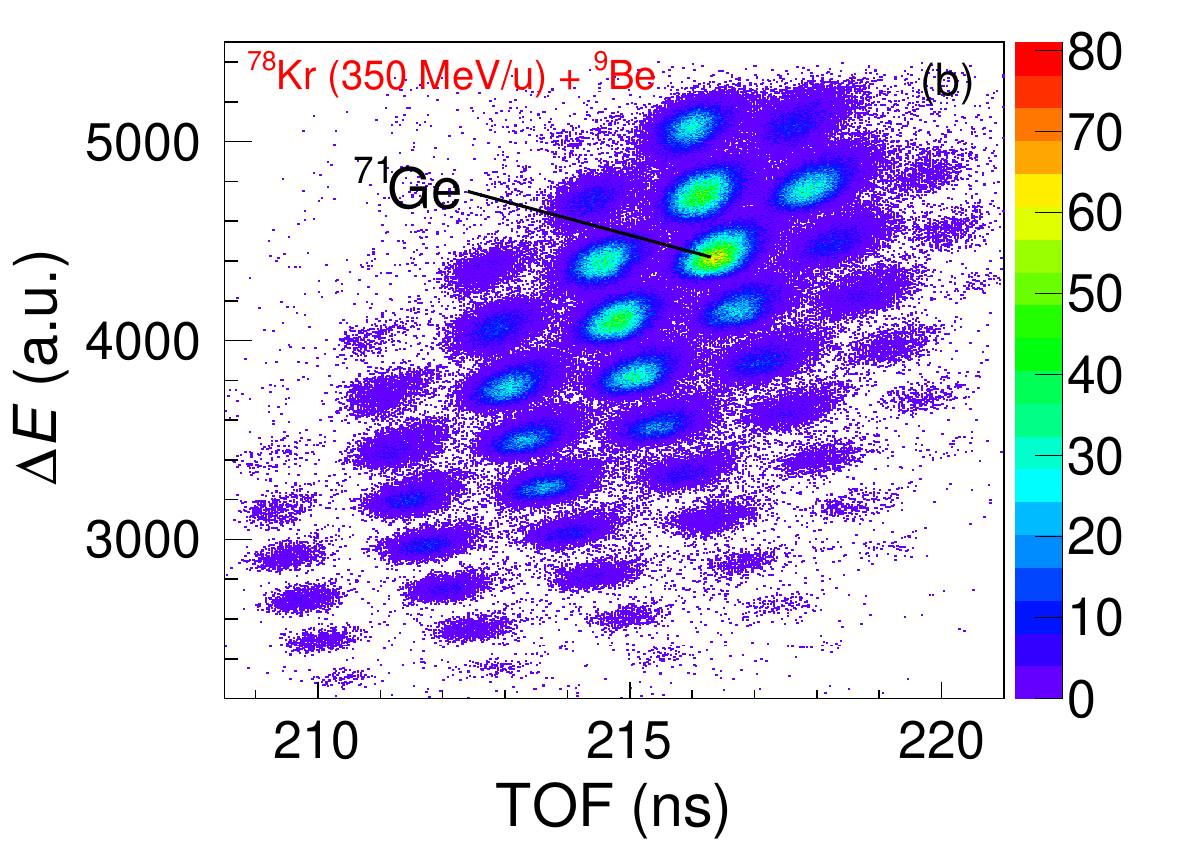}
\end{minipage}
\caption{Typical particle identification plots for the ions produced by the fragmentation of a 400 MeV/u $^{40}$Ar beam (a) and a 350 MeV/u $^{78}$Kr beam (b).~The abscissa shows the ion’s time-of-flight from F1 to F4 with the correction of ion's momentum using the ion's position measured at F3, while the ordinate displays the ion’s energy loss in MUSIC1. In both panels, the nucleus with largest intensity is indicated. It is worth mentioning that no energy degraders were used at the momentum-dispersive foci F1 and F3 for isotope separation.}
\label{fig2}
\end{figure}
%-------------------------------------------------------------------------------

The experiment was conducted using four different settings of the RIBLL2. In each setting, the ions of interest were separated and transported to the F4 area for CCCS measurements. In total, several tens of isotopes located from near the $N=Z$ line towards the neutron-rich region were produced with reasonable statistics. Taking Si isotopic chain as an example, various isotopes (from $N=Z-1$ nucleus $^{27}$Si to neutron-rich nucleus $^{34}$Si) were created. The particle identification can be achieved by utilizing the so-called $\Delta{E}$-B$\rho$-TOF method. For illustration, Fig.~\ref{fig2}(a) depicts a two-dimensional plot of ion’s $\Delta{E}$ versus their TOF for one of the RIBLL2 settings, wherein the $^{35}$P ions exhibit the largest intensity.~The slits at F1 were fully open during this setting.~The assignment of isotopes was realized using a specific method. During the experiment, we tuned the central B$\rho$ of RIBLL2 in several steps to gradually shift the coverage of isotopes from including primary beam species to consisting of the nuclei of interest. We chose a relatively small step size to ensure that there are some overlapping isotopes between consecutive steps. This approach allowed us to identify the ion species by referencing primary beam species’ location in the PID plot and the connection between adjacent settings. As clearly seen in Fig.~\ref{fig2}(a), each nuclide occupies a unique position in this PID plot and different ion species are well separated. The charge resolution exhibits a clear dependence on $Z$, varying from 0.12 ($\sigma$) to 0.18 ($\sigma$).~Regarding the ion's TOF, we conducted a phenomenological correction of the ion's momentum using the ion’s position at F3 measured by MWPC2 with a resolution ($\sigma$) of around 0.5 mm.~After the correction, the resolution of TOF was significantly improved, making it sufficient for distinguishing neighboring isotopes in all isotopic chains produced during the experiment.~Such good ability for PID ensures a successful measurement of CCCS. Recently, the CCCS of $^{28}$Si on carbon at 300 MeV/u has been deduced by analyzing the data from this experiment and the measured CCCS is consistent with the existing data at similar energies~\cite{Wang2023}.% by using the radiation sources,

To further test the RIBLL2's capability for separation and identification of secondary beams produced by the fragmentation of heavier projectiles, we carried out another experiment by employing a 350 MeV/u $^{78}$Kr beam.~After impinged on a 10 mm thick Be target, a cocktail beam including dozens of ion species were produced.~Figure~\ref{fig2}(b) is the PID plot for those ions reached F4 in one of the RIBLL2 settings. In this setting, the F1 slits were adjusted to have a momentum acceptance of $\pm$0.4\%.~It is evident that the fragmentation residues of $^{78}$Kr ions can be clearly identified, which demonstrate that RIBLL2 is now capable for separation and identification of medium-mass nuclei.

In summary, we constructed a new experimental platform at the last focal plane area of the RIBLL2 beam line in the HIRFL-CSR accelerator facility. Meanwhile, we developed several novel detectors and upgraded some existing ones.~To test the performance, we conducted two experiments at the RIBLL2-F4 platform by utilizing a 400 MeV/u $^{40}$Ar beam and a 350 MeV/u $^{78}$Kr beam, respectively. The first results reveal a clear separation in PID plots and an unambiguous identification for the secondary beam species produced by projectile fragmentation reactions, which demonstrate that the RIBLL2 separator is able to provide medium-mass RIBs and the newly-developed RIBLL2-F4 platform is fully operational. With the full realization, the RIBLL2 can be employed to produce RIBs of interest and then transport them to one of experimental terminals (i.e., ETF and RIBLL2-F4 platform) for conducting RIB experiments, and, naturally, can be used as a transfer beam line to better deliver beams from CSRm to CSRe for storage ring experiments.

It is worth noting that the next-generation RIB facility High Intensity heavy-ion Accelerator Facility (HIAF)~\cite{Yang2013,Zhou2022} is currently under construction at Huizhou, China. At HIAF, the high-energy radioactive ion beam line HFRS~\cite{Sheng2020,Sheng2024} will be a powerful fragment separator for the production of RIBs.~The RIBLL2 can be effectively used to test and evaluate the performance of beam line detectors developed for the HFRS. Furthermore, the valuable experiences gained from experiments conducted at the RIBLL2 can be applied to enhance the future operation of the HFRS at HIAF.

%============================
\textbf{Conflict of interest}\\
%============================
The authors declare that they have no conflict of interests.

%========================
\textbf{Acknowledgments}\\
%========================
The authors thank the staff in the accelerator division of Institute of Modern Physics for providing stable beams.~This work was supported partially by the National Natural Science Foundation of China (Grant Nos. 12325506, 11922501, U1832211, 11961141004, and 11905260), the Western Light Project of the Chinese Academy of Sciences, and the open research project of CAS large research infrastructures.

%============================
\textbf{Author contributions}\\
%============================
Zhi-Yu Sun, Yong Zheng, and Bao-Hua Sun supervised the project. Bao-Hua Sun, Yong Zheng, and Xiao-Dong Xu proposed and led the experiments. Xiao-Dong Xu and Yu-Nan Song processed the data. Xiao-Dong Xu wrote the original manuscript with contributions from all co-authors, while Bao-Hua Sun and Satoru Terashima were deeply involved in the discussion process. All authors participated in the experiments and manuscript revision.

%===============================================================================

%===============================================================================
\end{document}